\documentclass[a4paper,10pt]{JHEP3}
\usepackage{graphicx}
\def\dd{\displaystyle}
\title{\bf The Radiative Corrections to the Mass of the Kink Using an Alternative Renormalization Program}
\author{S. S. Gousheh$^{1}$ \footnote{ss-gousheh@sbu.ac.ir}, A. Mohammadi$^{1}$, M. Asghari$^{1}$, R. Moazzemi$^{2}$\footnote{r.moazzemi@qom.ac.ir} and F. Charmchi$^{1}$\footnote{f$\_$charmchi@sbu.ac.ir}\\ $^{1}$Department of Physics, Shahid Beheshti University G. C., Evin, Tehran
19839, Iran\\ $^{2}$Department of Physics, The University of Qom, Ghadir Blv., Qom 371614-611, Iran}
\maketitle
\abstract{
In this paper we compute the radiative correction to the mass of 
the kink in $\phi^4$ theory in 1+1 dimensions, using an alternative 
renormalization program. In this newly proposed renormalization 
program the breaking of the translational invariance and the topological 
nature of the problem, due to the presence of the kink, is automatically 
taken into account. This will naturally lead to uniquely defined position 
dependent counterterms.
We use the mode number cutoff in conjunction with the above program to 
compute the mass of the kink up to and including the next to the leading 
order quantum correction. We discuss the differences between the results
of this procedure and the previously reported ones.
}
\begin{document}
\section{Introduction}
The quantum radiative corrections to the mass of the solitons have
been of great interest since the 1970's, and has had a long,
complicated, and at times controversial history. In 1974, Dashen
\emph{et al} \cite{dashen} computed the one-loop correction to the
mass of the bosonic kink in $\phi^4$ field theory for the first
time. In that article they used the mode number cutoff method with
a continuous form for the phase shifts of the scattering states.
After Dashen, several other authors have used similar methods to
compute similar corrections for analogous problems
\cite{gervais,raja}. Ever since Dashen's work, several other
methods have been invented or used for analogous problems, the
most important five of which are the following. First, the energy
momentum cutoff using discontinuous form of the scattering states
phase shifts \cite{raja,reb97,goldhaber}. The above two approaches have
sparked remarkable  controversies \cite{flor1,reb2003,flor2}.
Second, the derivative expansion of the effective action using
summation of the series for the exactly solvable cases, which
embeds an analytic continuation, or Pad\'{e} approximation or
Borel summation formula for the approximately solvable cases
\cite{chan}. Third, the scattering phase shift method in which the
change in the density of states due to the presence of the
disturbance is represented by the scattering phase shifts
\cite{graham1}. Fourth, the dimensional regularization technique
in which the zero point energy of the free vacuum is subtracted
and dimensional regularization is used \cite{reb2,reb2002}. Fifth,
the zeta function regularization technique which completely
bypasses the explicit subtraction of free vacuum energy
\cite{bordag}. Obviously all these methods have eventually
confirmed the DHN result. As a side note we should mention that
analogous corrections to the mass of the bosonic kink have been
computed in supersymmetric models (see for example
\cite{reb2,nastase,susy,shif,rebhan2004}).

The presence of either non-trivial boundary conditions or
non- perturbative backgrounds, e.g. solitons, have
important manifestations in the physical properties of the
systems. In particular, an alternative renormalization program has been
proposed \cite{gousheh-1} which is fully consistent with the boundary conditions
and its use has led to new results for the Next to Leading Order (NLO) 
Casimir effect within $\phi^4$ theory
\cite{paper11,paper12}. The main purpose of this paper is to explore another 
manifestation of this newly proposed renormalization program by presenting  
an analogous study for systems with non-trivial backgrounds. In particular, we
calculate the quantum correction to the mass of the $\phi^4$ kink,
which is analogous to the Casimir problem with the kink as its static background.

In this paper we explain briefly the alternative renormalization 
program as tailored to our problem. The starting part of
our computation parallels closely Dashen's work. That is, we use
the mode number cutoff with continuous phase shifts. However, the
counterterms that we derive differ form the free counterterms, by 
which we mean the ones derived specifically for the free case, i.e. 
cases with no non-trivial boundary conditions or spatial backgrounds. 
The main issue of the alternative renormalization program is that the presence
of non-trivial boundary conditions or strong backgrounds such as solitons, 
which could also affect the boundary conditions, are in principle 
non-perturbative effects.
Therefore, they define the overall structure and the properties of
the theory and obviously cannot be ignored or even taken into
account perturbatively. 
The alternative renormalization program is founded on the principle
that the solution to the problem should be self-contained and the 
renormalization procedure be done self-consistently with the nature 
of the problem.
 
An additional justification supporting this method is the
fact that the presence of either non-trivial boundary conditions or
non-trivial backgrounds or both break the translational symmetry of
the system. In our case this occurs when we fix the position of
the soliton. Obviously the breaking of the translational symmetry
has many manifestations. Most importantly all the \emph{n}-point
functions of the theory will have in general non-trivial position
dependence in the coordinate representation. The procedure to
deduce the counterterms from the \emph{n}-point functions in a
renormalized perturbation theory is standard and has been
available for over half a century. This, as we shall show, 
could lead to uniquely defined position dependent
counterterms. Then, the radiative corrections to all the
input parameters of the theory, will be in general position
dependent. In that case, the information about the
non-trivial boundary conditions or position dependent background is
carried by the full set of \emph{n}-point functions, the resulting
counterterms, and the renormalized parameters of the theory. When
we compute the mass counterterm systematically by setting the
tadpole diagrams equal to zero, it turns out to be proportional to
Green's function,  as usual, which obviously has non-trivial
position dependence in this problem. This counterterm turns out to 
be different from the trivial sector only by some finite localized 
contributions which are proportional to the bound state distributions.

We have organized the paper in four sections as follows. In
Section \ref{sec.2} we set up the usual problem of $\phi^4$ theory
for a real scalar field in 1+1 dimensions, in the spontaneously
broken phase. We find the static background solutions which
include the trivial and the kink sectors. We also exhibit the
quantum fluctuations in both sectors, the latter of which includes
two bound states. 
Our main calculational tools in Section  \ref{sec3} are the renormalized 
perturbation theory and the expansion of the Lagrangian about the two 
different static sectors. 
We then calculate NLO correction to the kink mass by
subtracting the vacuum energies of the two sectors. The part of
this energy which does not depend on the counterterms is
calculated using the mode number cutoff. We then calculate the
contribution from the mass counterterms, which are fixed by the no
tadpole renormalization condition. When we add up all the
contributions we find an extra term which is due to our non-trivial
counterterm in the kink sector. Finally, in Section \ref{sec4} we
compare our methods and results to some earlier work.
\section{Kink solutions and their quantum fluctuations}
\label{sec.2}

\begin{figure}[htp]
\hspace{1cm}
\includegraphics[width=6cm]{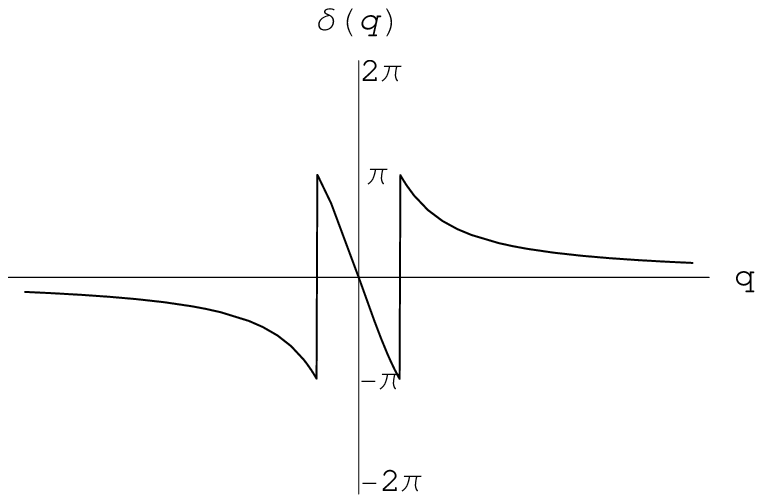}
\hspace{1cm}
\includegraphics[width=6cm]{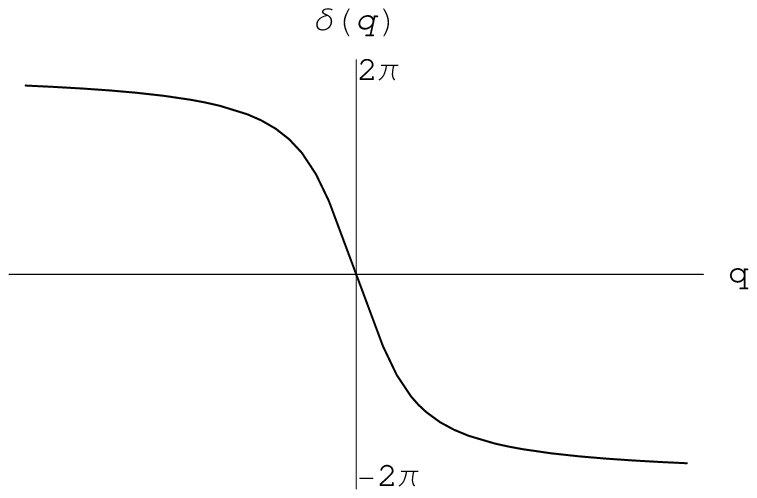}
\caption{\label{fig:1} \small A direct calculation of the phase
shift yields the graph on the top. However, in all physical
applications that we are aware of, the proper form to use is a
continuous as the one illustrated on the bottom.}
\end{figure}

In this section we shall very briefly state the standard results
for the static background solutions and their quantum fluctuations
for the bosonic $\phi^4$ theory. For a comprehensive review of the
standard materials, see for example \cite{raja}. We start with the 
Lagrangian density for a neutral massive scalar field, within 
$\phi^4$ theory, appropriate for the spontaneously broken symmetry 
phase in 1+1 dimensions,
\begin{equation}\label{1}
{\cal L} =\frac{1}{2}\left(\frac{{\partial \phi }}{{\partial
t}}\right)^2 - \frac{1}{2}\left(\frac{{\partial \phi }}{{\partial
x}}\right)^2 -U[\phi(x)],
\end{equation}
where $U[\phi]=\frac{\lambda_ 0}{4}\left( \phi^2 - \frac{\mu_
0^2}{\lambda_0} \right)^2$. The Euler-Lagrange equation can be
easily obtained and is a second-order non-linear PDE with the
following solutions: Two non-topological static solutions
$\phi_{\rm{vac.}}(x)=\pm\frac{\mu_ 0}{\sqrt \lambda_ 0}$, and two
topological static ones $\phi_{\rm{kink}}(x)=\pm\frac{\mu_ 0}{\sqrt
\lambda_ 0}\tanh [\frac{\mu_0(x-x_0)}{\sqrt 2}]$ which are called kink and
antikink, respectively. The presence of $x_0$ indicates the
translational invariance, and this will lead to a zero mode. The
total kink energy, sometimes called the classical kink mass can be
easily calculated and is given by $M_{\rm{cl.}}= \frac{{2\sqrt
 2 }}{3}\frac{{\mu_ 0^3 }}{\lambda_0   }$.
In order to find the quantum corrections to this mass, we have to
make a functional Taylor expansion of the potential about the
static solutions which yields the stability equation
\begin{equation}\label{3}
\left[ - \nabla ^2  + \frac{{d^2 U}}{{d\phi ^2 }}\bigg |_{\phi
_{\rm{static}} (x)} \right]\eta (x) = \omega'^2 \eta (x),
\end{equation}
where we have defined $\phi= \phi_{\rm{static}}+\eta$. The results
in the trivial sector are the following continuum states $ \eta(x)
= \exp (ikx)$ with $ \omega'^{2} = k ^2 + 2\mu_0^2$. In the kink sector we have the
following two bound states and continuum states:
\begin{eqnarray}
 \eta _0 (z) &=& \sqrt {\frac{{3m_0}}{8}} \frac{1}{{\cosh^2 z}},\nonumber\\
 \eta _B (z) &=&\sqrt {\frac{{3m_0}}{4}} \frac{{\sinh z}}{\cosh^2 z},\nonumber \\
 \eta _q (z) &=&\frac{{e^{iqz} }}{{N_q }}\left[-3\tanh^2 z+1+q^2+3iq\tanh z\right],
 \end{eqnarray}
where $m_0=\frac{\mu_0}{\sqrt{2}}$, $\omega_0^2=0$, $\omega_B^2=\frac{3}{4}m_0^2$ and 
$\omega_q^2=m_0^2(\frac{q^2}{4}+1)$. 
Here $N_q^2= 16\frac{{\omega_q^2}}{{m_0^4}}(\omega_q^2-\omega_B ^2 )$,
and $z=\frac{m_0x}{2}$. The continuum states $\eta _q(z)$ have the
following asymptotic behavior for $x \to \pm \infty$,
\begin{equation}\label{12}
 \eta _q (z) \to \exp [iqz \pm \frac{i}{2}\delta
(q)],
\end{equation}
where $\delta(q)=\dd-2\arctan[\frac{{3q}}{{2-q^2}}]$ is the
phase shift for the scattering states. We believe the phase shifts
should be in principle defined to be continuous functions of their
arguments. This is particularly apparent in their use in the
strong and the usual forms of the Levinson theorem 
(see for example \cite{gousheh-2}). For this
particular case the phase shift is illustrated in figure
\ref{fig:1}.

\section{First order radiative correction to the kink
mass}\label{sec3}

In this section we calculate the first order quantum correction to
the kink mass. As is well known, this is analogous to the Casimir problem for this case. That is,
the exact kink mass is the difference between the vacuum energies
in the presence and absence of the kink. To calculate this effect
we set up renormalized perturbation theory. We should mention that
in these (1+1)-dimensional problems, one usually chooses a minimal
renormalization scheme defined at all loops by
\cite{dashen,raja,reb97}
\begin{equation}
Z_\lambda=1\quad,\quad Z_\eta=1\quad\mbox{and}\quad m_0^2=m^2-\delta_m.
\end{equation}
 The sufficiency of these conditions is
supported by the fact that for any theory of a scalar field in two
dimensions with non-derivative interactions, all divergences that
occur in any order of perturbation theory can be removed by
normal-ordering the Hamiltonian \cite{coleman}. However, relaxing
the first condition might lead to some extra finite contributions, 
and this deserves a further investigation. Here, for mere
comparison reasons, we stay focused on the renormalization
program with the above stated conditions but with non-trivial mass
counterterm $\delta_{m_{\rm{kink}}}$.

Now we can split the expression for mass of the kink into the
following two parts,
\begin{eqnarray}\label{casimir}
M =(E_{\rm{kink}}-E_{\rm{vac.}})+(\Delta E_{\rm{kink}}-\Delta E_{\rm{vac.}}),
\end{eqnarray}
where the first part is in the vacuum sector of each, and the
second part is due to the counterterms. We put our solutions in a
box of length $L$ and impose periodic boundary conditions. The
continuum limit is reached by taking $L$ to infinity and the sum
turns into an integral. Now we use the usual mode number cutoff as
advocated by R.F. Dashen \cite{dashen} to calculate the first part
of Eq.\,(\ref{casimir}). In this method one subtracts the energies
of the bound states in the presence of solitons from the same
number of lowest lying quasi-continuum states in the vacuum of the
trivial sector. Then one subtracts the remaining quasi-continuum
states from each other in ascending order. In this case we have two 
bound states which are to be subtracted from $\omega'_{\pm1}$.
Then we subtract the quasi-continuum $q_n$ from the remaining
$k_{n+1}$ one by one. The periodic boundary condition implies,
\begin{equation}\label{18}
k_{n+1}L-2\pi=2n\pi=q_n \frac{mL}{2}+{\delta (q_n )}.
\end{equation}
The first part of Eq.\,(\ref{casimir}) can be easily calculated as
follows
\begin{eqnarray}
E_{\rm{kink}}-E_{\rm{vac.}}&&=M_{\rm{cl.}}+\frac{1}{2}(\sum\omega-\sum\omega')\nonumber\\
&&=\frac{m^3}{3\lambda}+\frac{1}{2}\left[\omega_{0}+\omega_{B}-
(\omega'_1+\omega'_{-1})+2\sum\limits_{n=1}^N {(\omega_n-\omega'_{n + 1})}\right]\nonumber\\
&&=\frac{m^3}{3\lambda}+ \frac{{\sqrt3 m}}{4}-m+\sum\limits_{n = 1}^N{\left[{m(\frac{q_n ^2}{4}
+1)^{\frac{1}{2}}-(k_{n+1}^2+ m^2)^{\frac{1}2}}\right]}\nonumber\\
&&\to\frac{m^3}{3\lambda}+\frac{{\sqrt 3 m}}{{4 }}-\frac{{3m}}{2\pi}
-\frac{{3 m}}{{4\pi}}\int_{-\infty}^\infty{\frac{{k^2+m^2/2 }}{{\sqrt{k^2+m^2}(k^2+\frac{{m^2 }}{4})}}}
\textrm{d}k,\nonumber\\\label{17}
\end{eqnarray}
where in the last step we have taken the continuum limit,
performed an integration by parts taking the appropriate boundary
values of the phase shift into account, as explained earlier. Note
that the last term is logarithmically divergent.

Now we calculate the second part of Eq.\,(\ref{casimir}):
\begin{eqnarray}\label{24}
\Delta E_{\rm{kink}}-\Delta E_{\rm{vac.}}=-\frac{1}{2}\int{\textrm{d}x\left[
\delta_{m_{\rm{kink}}}\phi^2_{\rm{kink}}(x)-\delta_{m_{\rm{vac.}}}\phi^2_{\rm{vac.}}(x) \right]},
\end{eqnarray}
where $\delta_{m_{\rm{kink}}}$ and $\delta_{m_{\rm{vac.}}}$ 
are the mass counterterms in the kink and vacuum backgrounds, 
respectively, and are calculated below. We first start with the mass 
counterterms in vacuum background. The procedure for obtaining 
this quantity is well known, e.g. by setting the tadpole equal to zero 
\cite{peskin}. The result is
\begin{equation}\label{26}
\delta_{{m}_{\rm{vac.}}}=\frac{{3\lambda }}{{4\pi}}\int_{-\infty}^\infty\frac{{\textrm{d}k} }{{\sqrt{k^2+m^2}}},
\end{equation}
which is logarithmically divergent.

Next we calculate the mass counterterm in the kink sector by
expanding the Lagrangian, which includes the mass counterterm,
around the kink background as follows
\begin{equation}
\phi (x,t) \rightarrow \phi_{\textrm{cl}}(x)+\eta(x,t)
=\frac{m}{{\sqrt\lambda}}\tanh(\frac{m}{{\sqrt 2 }}x) + \eta (x,t),
\end{equation}
where $\phi_{\textrm{cl}}(x)$ can be any of the static solutions, for example the
kink solution as indicated above. Then the Lagrangian which
incudes the mass counterterm becomes,
\begin{eqnarray}\nonumber\label{39}
{\cal L}&&=\frac{1}{2}(\partial_\mu\phi)^2+\frac{1}{2}(m^2-\delta_m)\phi^2
-\frac{\lambda }{4}\phi^4-\frac{(m^2-\delta_m)^2}{4\lambda}\nonumber \\
&&=\frac{1}{2}(\partial_\mu\eta )^2+\frac{1}{2}(m^2-3\lambda\phi_{\textrm{cl}}^2 )\eta^2
-\lambda \phi_{\textrm{cl}}\eta ^3-\frac{\lambda}{4}\eta^4-
\delta_m\phi_{\textrm{cl}}\eta-\frac{1}{2}\delta_m\eta ^2 \nonumber\\
&&-\frac{1}{2}(\partial _\mu  \phi_{\textrm{cl}})^2+\frac{1}{2}(m^2-\delta_m)\phi_{\textrm{cl}}^2
-\frac{1}{4}\lambda\phi_{\textrm{cl}}^4-\frac{{(m^2-\delta_m)^2}}{{4\lambda}}
+(m^2\phi_{\textrm{cl}}-\lambda\phi_{\textrm{cl}}^3
+\partial_\mu\phi_{\textrm{cl}}\partial^\mu)\eta.
\end{eqnarray}
Note that the last term in the above equation which is
proportional to $\eta$ vanishes exactly after an integration by
parts and using the equation of motion. Therefore, the condition of
setting the tadpole equal to zero simply becomes
\begin{eqnarray}\label{f.rules}
\raisebox{-5mm}{\includegraphics[width=1.7cm]{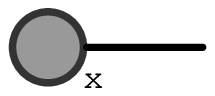}}\hspace{2mm}=
\hspace{2mm}\raisebox{-2mm}{\includegraphics[width=1.3cm]{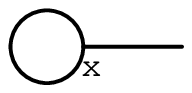}}\hspace{2mm}
+\hspace{2mm}\raisebox{-2mm}{\includegraphics[width=1.cm]{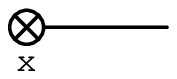}}\hspace{2mm}
+\dots=0.
\end{eqnarray}
Accordingly, up to first order in $\lambda$ we obtain,
\begin{equation}
i\delta_m(x,t)~\phi_{\textrm{cl}}(x)=\frac{1}{2}\bigg[(-6i\lambda)~\phi_{\textrm{cl}}(x)\bigg]G(x,t;x,t),
\end{equation}

where $G(x,t;x',t')$ is the propagator for the particular problem
under investigation. We finally obtain the following general
result, which is also obtained in \cite{paper11,paper12} using
analogous general arguments but a slightly different method,
\begin{equation}\label{43}
\delta_m(x,t)=-3\lambda G(x,t;x,t).
\end{equation}
Note that Eq.\,(\ref{26}) is a special case of this equation. More
importantly, note that the counterterms in general naturally turn
out to be position dependent. Since $G(x,t;x',t')$ is uniquely
determined by the nature of the problem, so is $\delta_m(x,t)$
via Eq.\,(\ref{43}). The Green function for this problem in the
presence of a kink is,
\begin{eqnarray}\label{28}
G(x,t;x',t')=i\int{\frac{{\textrm{d}\omega}}{{2\pi}}}
e^{i\omega(t-t')} \left(\sum\limits_{n \ne 0} {\frac{{\eta^*_n(x)\eta_n
(x')}}{{\omega^2_n-\omega^2}}}+\int{\textrm{d}k}
\frac{{\eta^*_k(x)\eta_k(x')}}{{\omega^2_k-\omega^2}}\right),
\end{eqnarray}
where the sum indicates the contributions of the bound states and
the integral the continuum states. Note that the zero mode is
neglected since it is only the manifestation of the translational
invariance of the system and is to be treated as a collective
coordinate \cite{raja,shif}. The above equation, when the two
space-time points are set to be equal and the $\omega$ integration
is performed, becomes
\begin{equation}\label{30.5} G(x,t;x,t)
=-\frac{{\eta_B^2(x)}}{{2\omega_B}}-\int{\frac{{\textrm{d}k}}{{2\pi}}} 
\frac{{\left|{\eta_k(k,x)}\right|^2}}{{2\omega_k}}.
\end{equation}
Calculating this integral is very cumbersome, but we can use an
interesting relationship which is the local version of the
completeness relation \cite{gousheh-3,gold,reb2002}:
\begin{equation}\label{31}
{\left|{\phi(k,x)}\right|^2=1-\frac{m}{{\omega_k^2-\omega_B^2}}
\eta_B^2(x)-\frac{{2m}}{{\omega_k^2}}\eta_0^2(x)}.
\end{equation}
Using the above equation, Green's function is easily computable by
performing simple integrals. Putting Eq.\,(\ref {31}) into Eq.\,(\ref{30.5}) 
and using Eq.\,(\ref{43}) the mass counterterm in the kink
background becomes,
\begin{equation}\label{32}
\delta_{m_{\rm{kink}}}=\frac{\lambda}{{\sqrt 3 m}}\eta_B^2
(x)-\frac{{3\lambda}}{{\pi m}}\eta_0^2(x)+\frac{{3\lambda}}{{4\pi }}
\int_{-\infty}^\infty {\textrm{d}k\frac{1}{{\sqrt{k^2+m^2}}}},
\end{equation}
which as expected earlier is different from mass counterterm in
the trivial sector, i.e. the last term in Eq.\,(\ref{32}). In fact
it has extra localized finite $x$-dependent terms due to the presence of the
bound states, and obviously this difference tends to zero as
$x\to\pm\infty$. An alternative reasoning is that the kink
solution also tends to either of the trivial vacuum states as
$x\to\pm\infty$. To complete the calculation we need to calculate
Eq.\,(\ref{24}) by inserting the expressions for $\delta_{m_{\rm{kink}}}$ 
and $\delta_{m_{\rm{vac.}}}$ into it. The result is
\begin{eqnarray}\label{33}
\Delta E_{\rm{kink}}-\Delta E_{\rm{vac.}}&=&\frac{-1}2\int_{-\infty}^\infty\textrm{d}x\bigg\{\delta_{m_{\rm{vac.}}}
\left[\phi^2_{\rm{kink}}(x)-\phi^2_{\rm{vac.}}(x)\right]\nonumber
+\left[\frac{\lambda }{{\sqrt 3 m}}\eta _B ^2 (x)-\frac{{3\lambda }}{{\pi m}}\eta^2_0(x)\right]
\phi^2_{\rm{kink}}(x)\bigg\}\\
&=&\dd\frac{m}{\lambda}\delta_{m_{\rm{vac.}}} -\frac{{\sqrt 3\pi-3}}{{20\pi }}m.
 \end{eqnarray}
Inserting the expressions obtained in Eqs.\,(\ref{17},\ref{33}) into Eq.\,(\ref{casimir}) we obtain the following expression for $M$:
\begin{eqnarray}\label{380}
M &&=\frac{{m^3 }}{{3\lambda }}+\frac{\sqrt 3 m}{4}-\frac{{3m}}{{2\pi}}
-\frac{{\sqrt 3\pi-3}}{{20\pi }}m \nonumber \\
&&-\frac{{3m}}{{4\pi}}\int\limits_0^\infty{\frac{{2k^2+m^2}}{{\sqrt{k^2+m^2}(k^2+\frac{{m^2}}{4})}}}
\textrm{d}k+ \frac{{3m}}{{2\pi }}\int\limits_0^\infty
{\frac{{dk}}{{\sqrt{k^2+m^2}}}}.
 \end{eqnarray}
The logarithmic divergences cancel and the final result is:
\begin{equation}\label{34}
M = \frac{{m^3 }}{{3\lambda }} + \frac{m}{{4\sqrt 3 }} -
\frac{{3m}}{{2\pi }} - \frac{{ \sqrt 3 \pi-3  }}{{20\pi }}m.
\end{equation}
Our result differs slightly from the previously reported result \cite{dashen}, 
by the last term in Eq.\,(\ref{34}).

\section{Conclusion}\label{sec4}
In this paper we have calculated the NLO correction to the mass of
the kink using the newly proposed alternative renormalization program. 
The use of this renormalization program is justified by the fact that 
the presence of non-trivial boundary
conditions or strong non-trivial backgrounds, such as solitons,
which could also affect the boundary conditions are in principle
non-perturbative effects. Therefore, they define the overall
structure and the properties of the theory and obviously cannot be
ignored or even taken into account perturbatively. We believe that
the solution to the problem should be self-contained and the
renormalization procedure be done self-consistently with the
nature of the problem. 
Moreover, the presence of a kink with a fixed position breaks
the translational symmetry of the system and this has profound
consequences. In particular all of the \emph{n}-point functions of
the theory, the counterterms and the renormalized parameters of
the theory will in general become position dependent. We have
shown this explicitly for the mass counterterm in this problem.
This will affect the quantum corrections to the kink
mass. In particular in Eq. (\ref{32}) we have shown
explicitly the difference between $ \delta_{m_{\rm{kink}}}$ and
$\delta_{m_{\rm{vac.}}}$. This has led to a small correction to
the result obtained by using free counterterms. 

\acknowledgments
It is a great pleasure for us to acknowledge the useful comments
of H.R. Sepangi. This research was supported by the research office 
of the Shahid Beheshti University.

\end{document}